%% file: main.tex
\documentclass[conference]{IEEEtran}
\IEEEoverridecommandlockouts
\usepackage{float}
\usepackage{cite}
\usepackage{gensymb}
\usepackage{caption}
\usepackage{subcaption}
\usepackage{geometry}
\usepackage{multicol}
\usepackage{lipsum}
\usepackage{slashbox}
\usepackage{tikz}
\usepackage[utf8]{inputenc}
\usepackage{amsmath,amssymb,amsfonts}
\usepackage{algorithmic}
\usepackage{graphicx}
\usepackage{textcomp}
\usepackage{xcolor}

\def\BibTeX{{\rm B\kern-.05em{\sc i\kern-.025em b}\kern-.08em
    T\kern-.1667em\lower.7ex\hbox{E}\kern-.125emX}}

\begin{document}

\title{Self Interference Management in In-Band Full-Duplex Systems\\}

\author{\IEEEauthorblockN{Hossein Mohammadi, Maryam Sabbaghian}
\IEEEauthorblockA{\textit{Dept. Electrical and Computer Engineering} \\
\textit{University of Tehran}\\
Tehran, Iran \\
\{Hossein.Mohammadi, msabbaghian\}@ut.ac.ir}
\and
\IEEEauthorblockN{Vuk MArojevic}
\IEEEauthorblockA{\textit{dept. Electrical and Computer Engineering} \\
\textit{Mississippi State University}\\
USA \\
vuk.marojevic@msstate.edu}
}

\maketitle
\pagenumbering{roman}
\begin{abstract}

    \label{sec:abst}
    \input{Sections/abstract.tex}
\end{abstract}

\section{Introduction}
\label{sec:intro}
\input{Sections/Introduction.tex}

\section{
Problem Formulation}
\label{sec:problem}
\input{Sections/problem.tex}

\section{
Beamforming vs. ANN-Based SIC}
\label{sec:sys model}
\input{Sections/sys_model.tex}

\section{Simulations}
\label{sec:sim result}
\input{Sections/sim_result.tex}

\section{conclusion}
\label{sec:conc}
\input{Sections/conclusion.tex}


\bibliographystyle{IEEEtran}
\bibliography{main}

\end{document}

%% file: Sections/abstract.tex
The evolution of wireless systems has led to a continuous increase in the demand for radio frequency spectrum. To address this issue, a technology that has received a lot of attention is In-Band Full-Duplex (IBFD). The interest in IBFD systems stems from its capability to simultaneously transmit and receive data in the same frequency. Cancelling the self interference (SI) from the transmitter to the collocated receiver plays a pivotal role in the performance of the system. 
There are two types of SI cancellation (SIC) approaches, passive and active. In this research, the focus is on active cancellation and, in particular, SIC in the digital domain. 
Among the direct and backscattered SI, the former has been studied for a long time; therefore, the backscatter is considered in this research and two SIC approaches are analyzed. The first achieves SIC through beamforming. This requires knowing the angle of the received SI 
to put the beam null-space in this direction. 
The second method removes SI by employing an Artificial Neural Networks (ANNs). Using an ANN, there is no need to know the direction of the SI. The neural network is trained with pilots which results in the network being able to separate the desired signal from the SI at the receiver. 
Bayesian Neural Networks 
show the importance of the weights and assign a parameter that facilitates ignoring the less significant ones. Through comparative simulations we demonstrate that the ANN-based SIC achieves equivalent bit error rate performance as two beamforming methods. 

\vspace{3mm}
\begin{IEEEkeywords}
IBFD, SIC, Beamforming, LCMV, MVDR, MIMO, ANN, Bayesian Neural Network.
\end{IEEEkeywords}

%% file: Sections/Introduction.tex
Next generation wireless communications are characterized by 
very high data rates (order of Gbps), extremely low latency, increased base station capacity and notable improvement in the users' perceived quality of service (QoS).\cite{1agiwal16}, \cite{112014scenarios5G}. Owing to the ever-increasing data rates which are supposed to be delivered by communication networks, the spectral efficiency of the networks needs to be further improved\cite{25dsouza2020symbol}. Although advanced communications techniques, such as multiple-input multiple-output (MIMO) and orthogonal frequency division multiplexing (OFDM), 
have been proposed as promising solutions for increasing the network's spectral efficiency, they are unable to fulfil the emerging service needs. 

Today's systems use either time division or frequency division duplex for transmitting and receiving signals. In other words, only half-duplex (HD) operations are employed in a given band. This lowers the resource utilization; i.e. in comparison with FD, HD reduces capacity 
by a factor of two \cite{3zhang2016full}. A promising solution for overcoming the limitation of HD is using IBFD systems; however, the huge gap between the transmitted and received signal power makes the use of IBFD challenging \cite{3zhang2016full}, \cite{6sabharwal2014}, \cite{7liu2015band}. Among the proposed solutions, beamforming is the most convincing one \cite{8zheng2014jointBeam} although it needs a precise estimation of the interference angle. This allows putting null spaces into the self interference (SI) angles \cite{9beamformingSpace}. In addition, among the pervasive algorithms such as Linear Constraint Minimum Variance (LCMV) and Minimum Variance Distortionless Response (MVDR), matrix inversion increases the complexity in particular for massive MIMO systems \cite{16Boroujeny}, \cite{17van2004optimum}.

\indent In order to address the aforementioned problems, this article proposes to use an artificial neuronal network (ANN) for the following reasons: The first and foremost reason is the nonlinear characteristic of the interference which can be properly modeled by a neural network \cite{12ANNFD2018non}, \cite{22zhang2018self}. References \cite{13digitalahmed2015all}, \cite{14analog2011practical}, \cite{15analog2008zigzag} propose a combination of SI cancellation (SIC) in the analog and digital domains with the aim of suppressing the SI to the level of the noise floor; however, perfectly cancelling the SI in the analog domain is highly challenging and costly \cite{24mohammadi2022ai}. We therefore study digital cancellation using an ANN. 
\\
\indent Despite the potential benefits of employing an ANN for this purpose, there are known complications. 
As discussed in detail in \cite{4ANNHaykin}, \cite{5Bishop} and \cite{23mohammadi2021artificial}, it is impossible to assure that the obtained ANN weights are optimal since the cost function is non-convex; therefore, based on the initial values of the weights, a solution may reach a global or a local minimum. In order to verify that the weights converge to appropriate values given a network structure, the proposed solution is compared with conventional algorithms. More precisely, we compare the output of the ANN with the LCMV and MVDR algorithms, which are commonly employed in full-duplex systems, and show that 
the proposed ANN learns correctly and is able to suppress the SI. 

The rest of this paper is as follows. Section II formulates the problem that we address in this paper. Section \ref{sec:sys model} describes the system model of the conventional SIC methods 
and the proposed ANN solution. In Section \ref{sec:sim result} simulation results are presented and discussed. Section \ref{sec:conc} derives the conclusion and future research needs. 

%% file: Sections/problem.tex
\indent Conventional wireless devices work in half-duplex mode on a single frequency channel in order to avoid the high-power transmit to leaking into 
the received signal. This is known as SI. 
It has been demonstrated that full-duplex wireless communication is possible by mitigating the SI through a combination of passive suppression and active cancellation techniques 
\cite{10everett2014passive}. 
It should be noted that the changing angle of the SI due to reflections in a dynamic environment, makes passive suppression techniques fail to remove the interference from the desired signal. 
Since the direct path interference has been studied extensively in literature this research focuses on the SI resulting from backscatter.  
This is illustrated in Fig. \ref{fig:SIModel}. 
We propose a new approach to SI cancellation using ANNs and compare it against the traditional beamforming based solution.

%% file: Sections/sys_model.tex
This section presents the two SIC approaches: beamforming and ANN. The former is discussed briefly with the aim of comparison for the latter. First we formulate the receiver structure and filter taps to overcome the SI using two general algorithms. Our proposed ANN structure is introduced as an alternative way to implement an equalizer in the receiver to enable full-duplex communications.

\subsection{Beamforming}
By means of beamforming we can also create nulls in the desired angle of arrival, in this case from the strong reflected signals. Two algorithms are considered: LCMV and MVDR. 
\begin{figure}[b]
    \centering
    \includegraphics[scale=0.8]{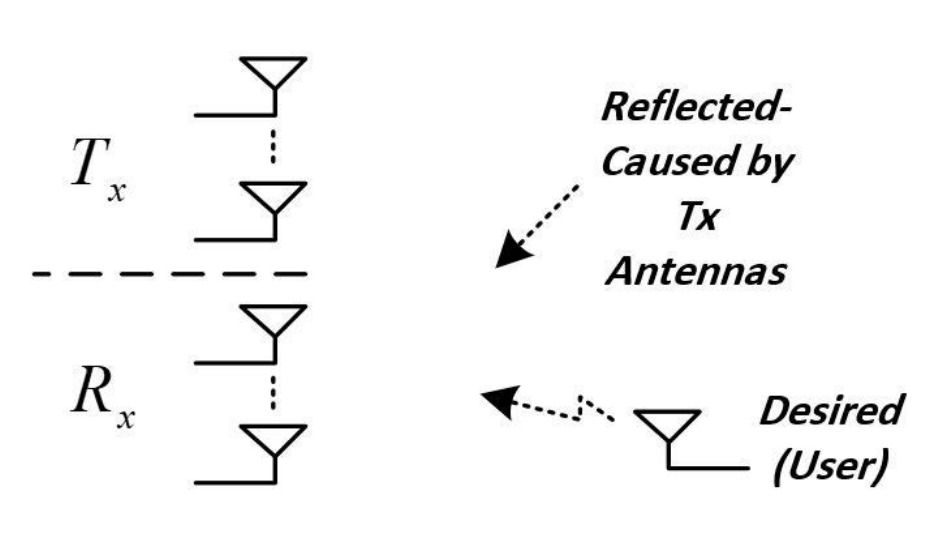}
    \caption{A simple model of the received SI and the desired signal.}
    \label{fig:SIModel}
\end{figure}
\subsubsection{MVDR}
\begin{figure}[t]
    \centering
    \includegraphics[scale=1.1]{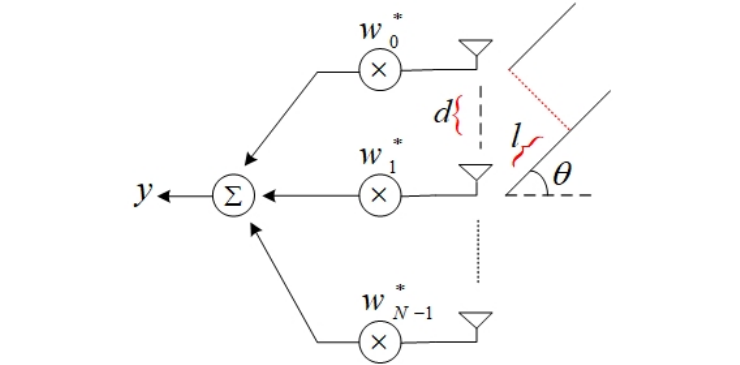}
    \caption{Model of the signal entering the receiver with N filter taps.}
    \label{fig:filtertaps}
\end{figure}

As shown in Fig. \ref{fig:filtertaps}, the MVDR algorithm finds the filter taps in order to minimize the power of the received signals except those coming from the desired direction. The received signal is then equal to
\begin{equation}
    Y_{in}(\omega) = Y_D(\omega) + Y_{Int}(\omega) + Y_\nu(\omega).
    \label{eq:eq1}
\end{equation}
\(Y_D(\omega)\) is the desired received signal, \(Y_{Int}(\omega)\) is the received interference and \(Y_\nu(\omega)\) is the noise in frequency domain. This signal passes through a filter whose output is
\begin{equation}
    Y_o(\omega) = W_{1 \times N}^H(\omega) Y_{in}(\omega).
    \label{eq:eq2}
\end{equation}
Adhering to the MVDR definition, we need to minimize interference and noise simultaneously and maximize the power of the desired signal. Therefore, we have:
\begin{equation}
\begin{split}
   & min \Big\{\big|Y_{\nu,i}(\omega) \big|^2 \Big\},\\
    & s.t.\ \underline{W}^H \underline{V}(k_d) = 1.
    \label{eq:eq3}
\end{split}
\end{equation}
\(\underline{W}^H\) is the filter taps vector and \(\underline{V}(k_d)\) is the steering vector in the desired direction. The optimum filter taps are obtained by using the Lagrange multiplier  \cite{16Boroujeny}, \cite{17van2004optimum} as
\begin{equation}
    \underline{W}_{MVDR}^H = \frac{\underline{V}^H S_{\nu}^{-1}}{\underline{V}^H {S_\nu}^{-1} \underline{V}}.
    \label{eq:eq4}
\end{equation}

\noindent Parameter \({s_\nu}^{-1}\) which is found by applying the matrix inversion lemma is equal to \cite{17van2004optimum}

\begin{equation}
    s_{\nu}^{-1} = \frac{1}{\sigma_{\nu}^2} \Big[I - \frac{\underline{V}_{int}^H \underline{V}_{int}}{S_I^{-1} + \frac{\underline{V}_{int} \underline{V}_{int}^H}{\sigma_{\nu}^2}}\Big].
    \label{eq:eq5}
\end{equation}

\subsubsection{LCMV}
As explained, the MVDR algorithm minimizes both noise and interference together by providing a compromise between them. In contrast to that, the LCMV algorithm puts a null space in the direction of the interference \cite{18souden2010study}. It is worth noting that the merit of LCMV shows itself when we have a strong interference in a specific direction \cite{17van2004optimum}. Hence, we have

\begin{equation}
\begin{split}
       & min \Big\{\big|Y_{\nu}(\omega) \big|^2 \Big\},\\
    & s.t.\ \underline{W}^H C = g^H,
    \label{eq:eq6}
    \end{split}
\end{equation}

\noindent where
\begin{equation}
    C = [\underline{V}(k_d)\ \underline{V}(k_{i1})\ ...\ \underline{V}(k_{iL})], \quad g^H = [1\ 0\ ...\ 0].
    \label{eq:eq7}
\end{equation}

We define \textit{R} as our received signal vector which contains the desired signal, interference and noise. We then compute

\begin{equation}
    A = R^HR,
    \label{eq:eq8}
\end{equation}

\noindent which we can decompose 
as follows:

\begin{equation}
    A = Q\Lambda Q^H.
    \label{eq:eq9}
\end{equation}
\textit{Q} contains the eigenvectors and \(\Lambda\) is a diagonal matrix of eigenvalues. Therefore, the eigenvectors matrix is as follows:

\begin{equation}
    C_{EVD} = \big [\, \underline{Q}_1\ \underline{Q}_2\ ...\ \underline{Q}_{N_m} \big].
    \label{eq:eq10}
\end{equation}

\(N_m\) can be found from the eigenvalues, since based on the power of the interfering and the desired signals, the eigenvalues decrease sharply. The constraint in (\ref{eq:eq6}) then changes to

\begin{equation}
    \underline{W}^H C_{EVD} = g^H.
    \label{eq:eq11}
\end{equation}

The reason for applying eiqenvalue decomposition stems from the fact that in some scenarios the received interference might be weak with respect to the desired signal; therefore, there is no need to put a null space in the angle of interference. This 
allows the user signal to be received even though it falls in the interference direction. Hence, the filter taps are obtained as \cite{17van2004optimum}, \cite{16Boroujeny}

\begin{equation}
    \underline{W}^H = \underline{g}^H (C_{EVD}^H\, S_{\nu}^{-1}\,C_{EVD})^{-1} C_{EVD}^H\, S_{\nu}^{-1}.
    \label{eq:eq12}
\end{equation}

\subsection{Artificial Neural Networks}

\begin{figure}[t]
    \centering
    \includegraphics[scale = 0.6]{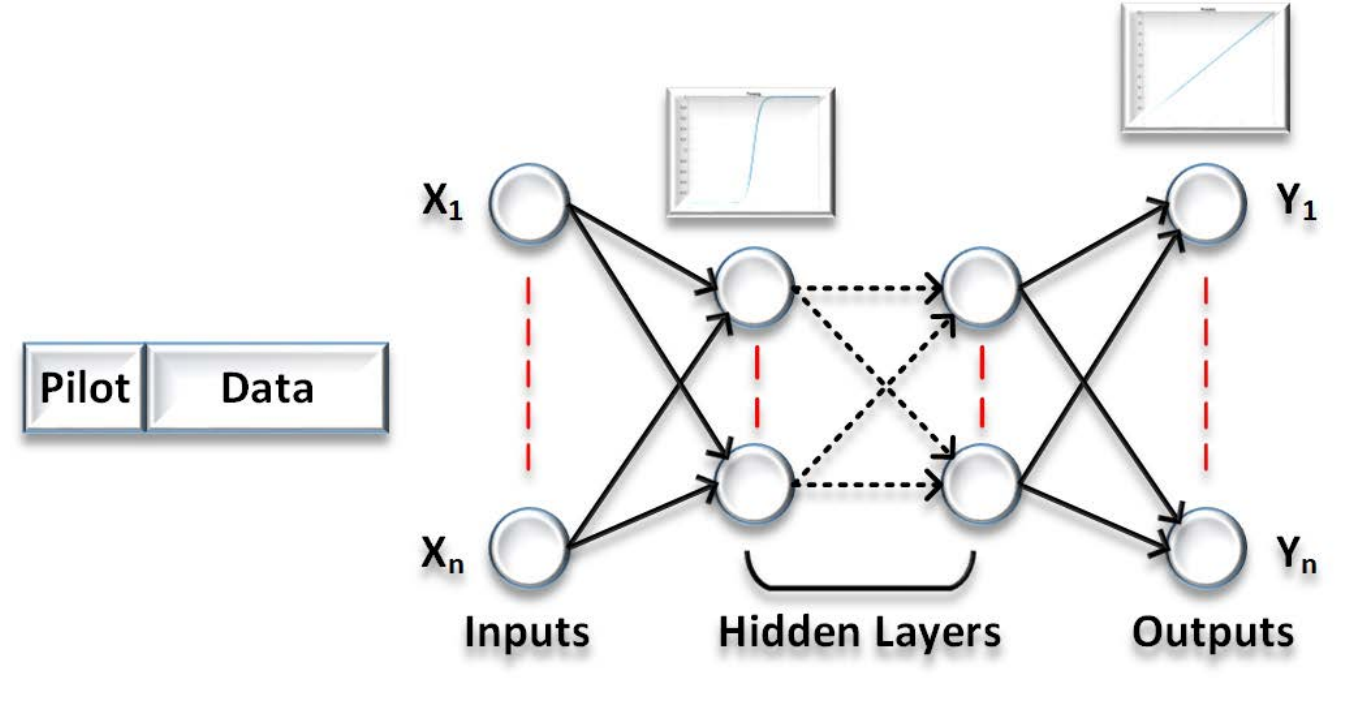}
    \caption{Proposed equalizer using an ANN with $M$ hidden layers.}
    \label{fig:annstruc}
\end{figure}

\begin{figure}[b]
    \centering
    \includegraphics[scale = 0.75]{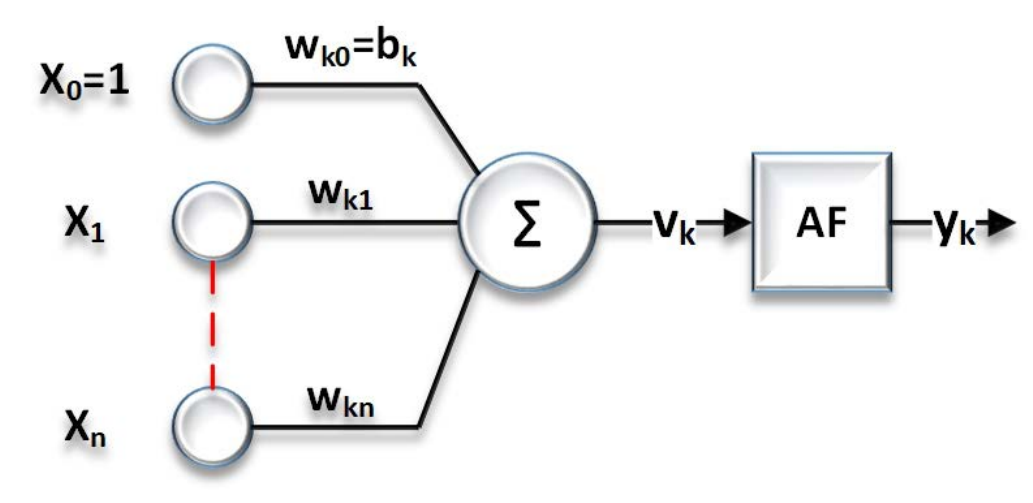}
    \caption{Structure of a single neuron.}
    \label{fig:ann1}
\end{figure}

ANNs have been proposed an alternative approach to mitigate the SI \cite{12ANNFD2018non}. The first and foremost reason is the nonlinear characteristic of the SI. References \cite{4ANNHaykin}, \cite{5Bishop} and 
\cite{yao2019ai} have shown that any nonlinear function can be modeled using an ANN. However, two distinct approaches exist for learning: supervised learning and unsupervised learning. 
We consider supervised learning, which requires labeled data 
to 
configure the neural network \cite{4ANNHaykin}. This can be achieved by pilot symbols in each specific block as shown in Fig. \ref{fig:annstruc}. 
The labeled data is divided into three distinct categories which are training, validation and test data. It is worth noting that the split is random, but usually more than half of the set is selected for training the ANN. 
The operation of a neuron with $n$ inputs is illustrated in Fig. \ref{fig:ann1} and is described as

\begin{equation}
    \begin{split}
        v_k &= \sum_{i=0}^{n} w_{ki}x_i \quad k = 1\ ... \ N\\
        y_k &= f(v_k).
    \end{split}
    \label{eq:eq13}
\end{equation}

\begin{figure}[t]
    \centering
    \includegraphics[scale = 0.22]{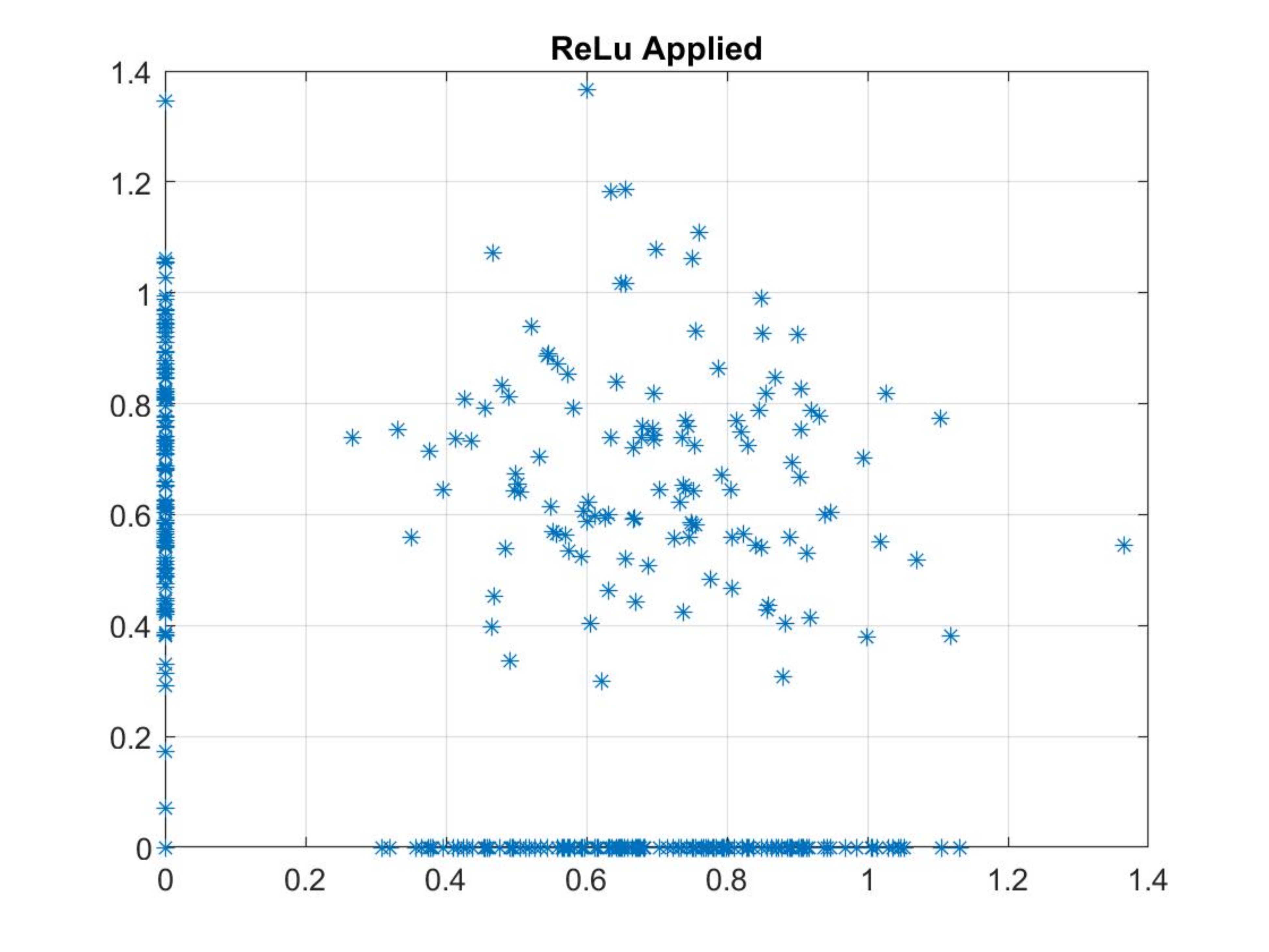}
    \caption{Mapping the symbols by using the ReLu activation function which causes the symbols in region two to be mapped to the y-axis, the symbols in region four to be mapped to the x-axis, and the symbols in region three to be mapped to the origin.}
    \label{fig:ReLu}
\end{figure}

Since the average power of the symbols is considered to be one, for the activation function, the sigmoid function,

\begin{equation}
    f(x) = \frac{2}{1+e^{-2x}} - 1,
    \label{eq:eq14}
\end{equation}
is considered in order to limit the output of each neuron to [-1,1].
The ReLu activation function, on the other hand, is known for its simplicity 
when compared to the sigmoid function. Fig. \ref{fig:ReLu} however shows that applying the ReLu activation function results in decreased distance among the symbols in the constellation diagram, degrading the BER.

\indent Using the error back propagation approach for learning the weights, the Bayesian Neural Network (BNN) is considered for its privileges over other algorithms such as the Levenberg Marquardt Algorithm (LMA). A BNN assumes a prior distribution for synaptic weights and applies a penalty for wrong decision. This significantly enhances the effectiveness of the learning process \cite{19burden2008bayesian}, \cite{5Bishop}.  shows, The proposed ANN is simulated for seven different channel scenarios. These are provided in Table \ref{tab:1}, which also shows the number of iterations and its best epoch for choosing the weights, number of effective weights in learning the network, the stopping factor and convergence time for each scenario.

%% file: Sections/sim_result.tex
\begin{table*}[t]
\caption{Comparison of different scenarios with their parameters}
\begin{center}
    
\begin{tabular}{|c||*{10}{c|}}\hline
\makebox[5em]{Scenario}
&\makebox[5em]{Angle}&\makebox[5em]{Power (dBW)}&\makebox[5em]{Iteration/Best}&\makebox[5em]{Weights}&\makebox[5em]{Stopping}&\makebox[5em]{Time (s)}\\\hline\hline
EPA &[60,20,80,-30] &[0,-1,-2,-3] &25/16 &48 &Minimum Gradient &10.28\\\hline
$1^{st}$ &[113,146,-134,149] &[-20,-30,-40,-50] &23/10 &48 &Minimum Gradient &5.5\\\hline
$2^{nd}$ &[113,146,-134,149] &[-8,-15,-20,-30] &22/8 &48 &Minimum Gradient &4.98\\\hline
$3^{rd}$ &[-95,-105,-130,-150] &[-4,-5,-7,-8] &17/11 &48 &Minimum Gradient &6.84\\\hline
$4^{th}$ &[-95,-105,-130,-150] &[-15,-18,-20,-25] &21/11 &48 &Minimum Gradient &5.75\\\hline
$5^{th}$ &[145,160,50,25] &[-15,-18,-20,-25] &25/8 &48 &Minimum Gradient &9.5\\\hline
$6^{th}$ &[-60,-85,-90,-115] &[-2,-5,-10,-12] &12/9 &48 &Minimum Gradient &3.31\\\hline
\end{tabular}
\label{tab:1}
\end{center}
\end{table*}
As described in the previous section, the LCMV and the MVDR algorithm are discussed with the aim of comparing their results in cancelling the SI with that of using an ANN. The communication scenario considered in this paper has 10 antennas at the receiver and one user in the far field transmitting the desired signal. Although 
direct SI path is considered as the worst case of interference leakage, our goal is to study the impact of the reflected SI paths for the proposed frequency selective channel. Therefore, the channel model is the Extended Pedestrian A (EPA) model with relative powers of [0 -1 -2 -3] dB. Other scenarios are also applied in order to check that the proposed system works properly. Table \ref{tab:1} captures the interference angles and the respective signal powers. 

The desired signal is located at 30$^{\circ}$ for all the scenarios. As a proof that our proposed ANN works in different scenarios, the number of iterations and the iteration which results in the best weights are provided. 
Table \ref{tab:1} also shows that for the EPA model and by considering the direct path we need the maximum number of iterations to get the best weights and for others even the $5^{th}$ and $6^{th}$ models in which we have correlation between the received antennas (changing the distance between antennas), the ANN achieves the best weights at almost the same iteration. To elaborate on this, since we employ a BNN, the number of effective weights and the stopping factor to learn is included the table. For all the scenarios the stopping factor is the minimum gradient. When it is very close to zero, it causes the ANN to stop learning. This means that we might be in a local or global minimum. However, the BER plots clearly show that the achieved ANN weights are indeed leading to the global minimum because they match the results of the optimum receiver structure, such as LCMV.
\begin{figure}[H]
    \centering
    \includegraphics[scale = 0.08]{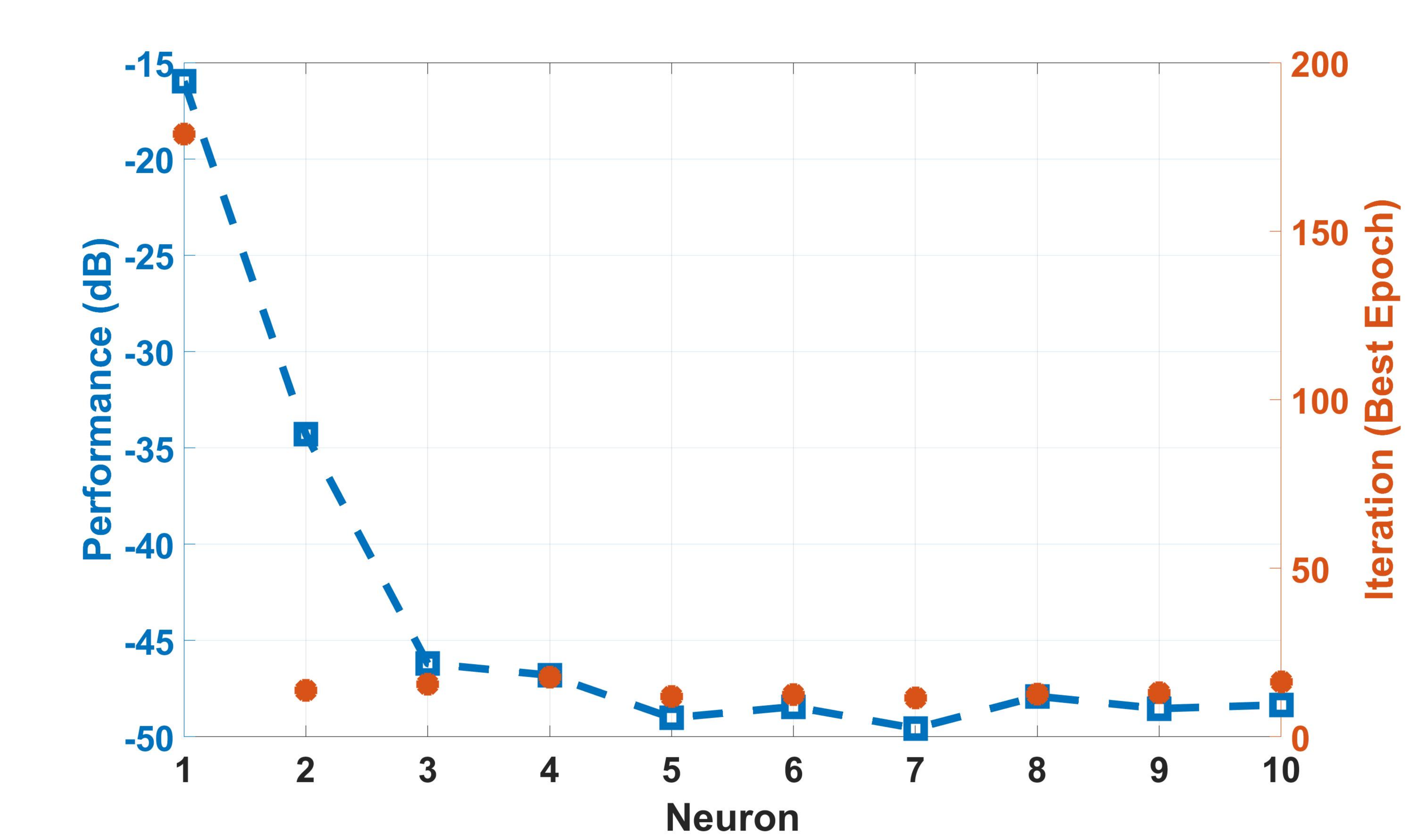}
    \caption{MSE and best epoch over the number of neurons.}
    \label{fig:neuron}
\end{figure}

\begin{figure}[h]
    \centering{\includegraphics[scale=0.08]{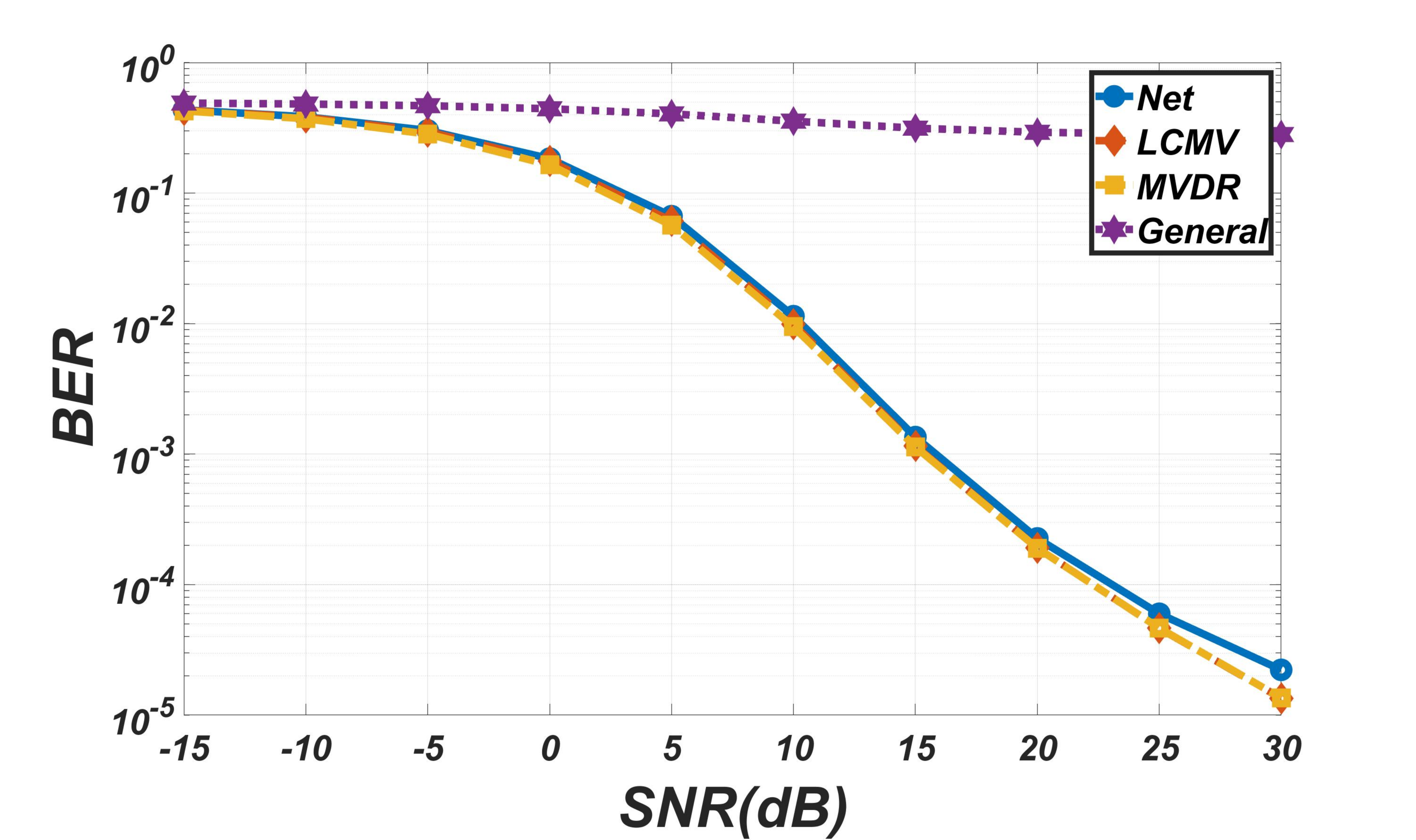}}
    \caption{BER comparison of ANN with LCMV and MVDR for EPA scenario.}
    \label{fig:BER}
\end{figure}

\begin{figure}[H]
\vspace{-5mm}
    \centering
    \includegraphics[scale = 0.08]{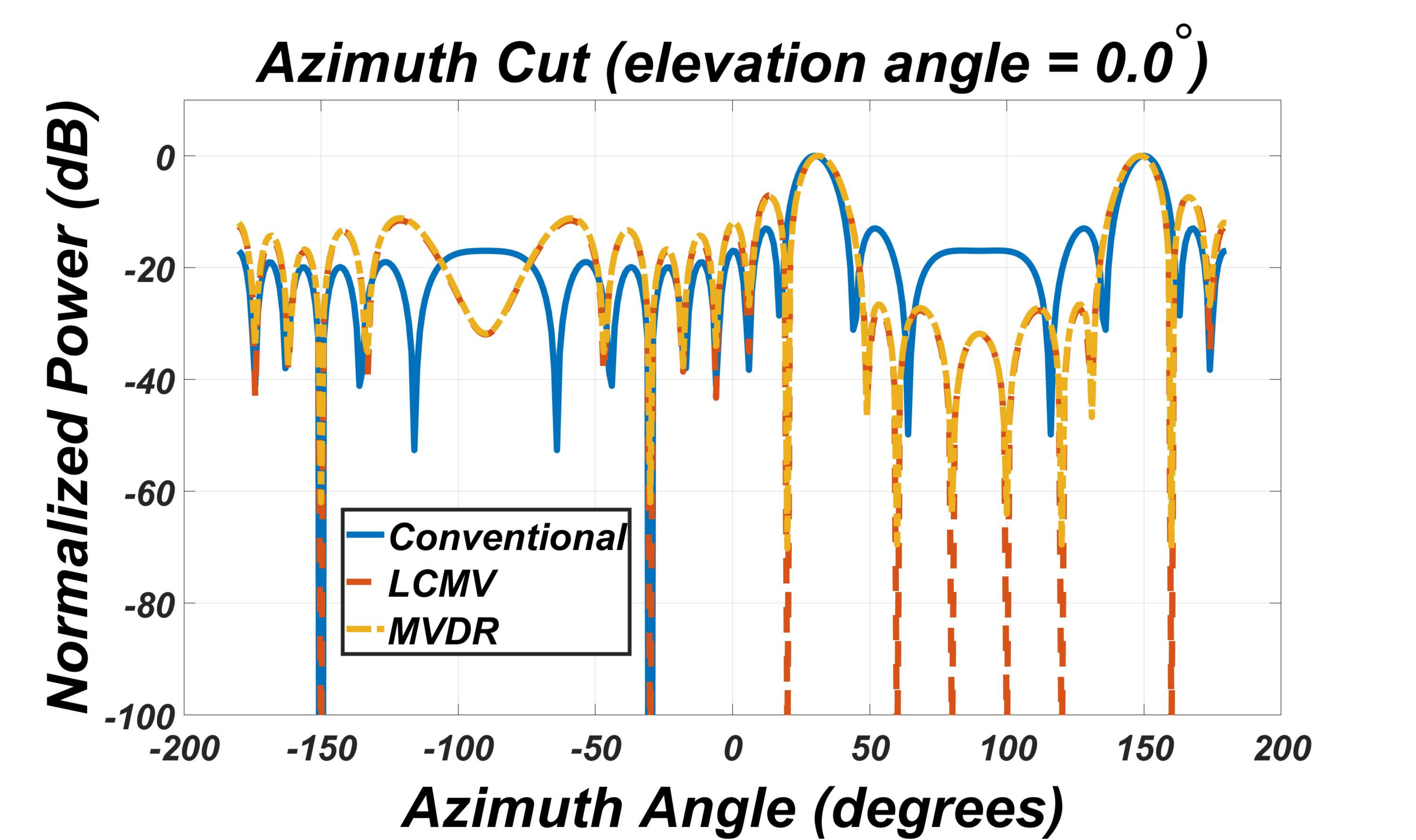}
    \caption{Removing interference with LCMV and MVDR for EPA scenario.}
    \label{fig:beam}
\end{figure}

\indent Fig. \ref{fig:neuron} plots the ANN performance over 
the number of neurons in the single hidden layer ($M$ = 1) as well as the best epoch for choosing the weights. The performance of the network reaches a steady state 
with $N$ = 3 neurons. We choose $N$ = 2 
for the sake of complexity and
to avoid over-fitting. 
Note that there are 20 inputs, being the in-phase and quadrature components for 10 antennas, and two output neurons, representing the complex sample to be demodulated. 

\begin{figure}
    \centering
    \includegraphics[scale=0.09]{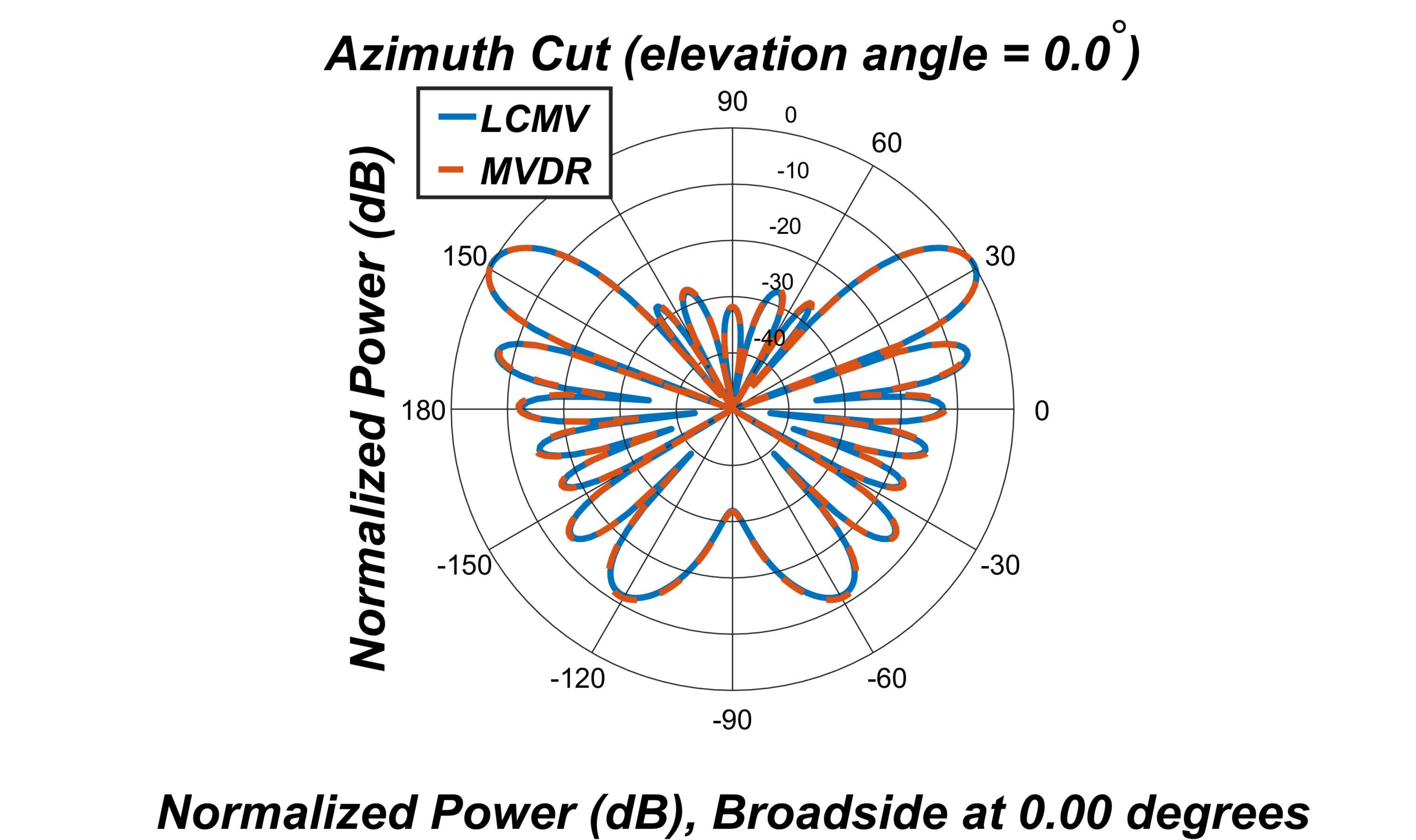}
    \caption{LCMV and MVDR interference cancellation for EPA scenario in polar coordinates. These two algorithms are removing interference clearly however we have a symmetry in nulling the interference owing to having the omnidirectional antennas.}
    \label{fig:beam_polar}
\end{figure}

\indent Table \ref{tab:1} presents the data for the different scenarios. Because of space limitations we illustrate the simulation results---BER and beam pattern---only for the EPA model, which features the direct path as the worst case. 
As Fig. \ref{fig:BER} shows, our proposed structure performs almost the same as the LCMV and the MVDR algorithms without prior knowledge of the interference angle. 
The ANN is able to extract the angle of the desired signal from the pilot symbols to remove the interference from it. This is the key advantage of using an ANN over other algorithms, including LCMV and MVDR.\\
\indent Since we are emphasizing on the interference removal, Figs.~\ref{fig:beam} and \ref{fig:beam_polar} illustrate the rectangular and polar plots of the LCMV and MVDR beam patterns for the EPA scenario. The conventional (capon) beamforming method is also simulated to show how these two algorithms minimize interference in the desired signal 
direction. The two reference algorithms null the interference at the angles of [60, 20, 80, -30] perfectly, as is evident from these figures. 
Note that the symmetry in nulling the interference is a result of using omnidirectional antennas.

%% file: Sections/conclusion.tex
\indent This paper addresses the pivotal problem of SI which is present in IBFD systems and introduces a new SI cancelling receiver using an ANN. We compare two distinct approaches, beamforming and ANN. For the beamforming approach, two common algorithms are employed to mitigate the SI. 
The main drawback of beamforming-based SIC is the need for having the exact knowledge of the interference angle and power spectral density of the noise. 
In contrast, the proposed ANN needs to dedicate only ten percent of the block length to pilot symbols and can extract all the necessary information for removing the interference from this data. Our results demonstrate that using an ANN is an effective alternative to traditional SIC approaches for enabling practical IBFD systems. 